\documentclass[epj]{webofc}
\usepackage[varg]{txfonts}   
\wocname{EPJ Web of Conferences}
\woctitle{CONF12}
\begin{document}
	\selectlanguage{english}
	\title{Analytic Solutions of Transverse Magneto-hydrodynamics under Bjorken Expansion}
	%
	%
	
	\author{Shi Pu\inst{1} \and
		Di-Lun Yang\inst{2}\fnsep\thanks{\email{dilunyang@gmail.com}} 
	}
	
	\institute{Institute for Theoretical Physics, Goethe University,
		Max-von-Laue-Str. 1, 60438 Frankfurt am Main, Germany.
		\and
		Theoretical Research Division, Nishina Center, RIKEN, Wako, Saitama 351-0198, Japan.
	}
	
	\abstract{%
		We review the recent developments of analytic solutions in transverse magneto-hydrodynamics under Bjorken expansion. It is found that the time dependence of magnetic fields can either increase or reduce the energy density depending on the decay exponent of magnetic fields. Moreover, perturbative solutions under weak magnetic fields with spatial inhomogeneity results in transverse flow, where the directions of flow also depend on the decay exponent of magnetic fields in time.    
	}
	\maketitle
	\section{Introduction}
	\label{intro}
Recently, the influence of strong magnetic/electric fields on the hot and dense matter such as the quark-gluon plasma (QGP) created in relativistic nucleus-nucleus collisions has been intensively studied. The fast-moving nuclei in peripheral collisions could produce extremely strong magnetic fields of the order of $B\sim10^{18}-10^{19}G$ in early times \cite{Deng:2012pc}. Such strong magnetic fields may influence the medium properties of QGP and lead to different effects including the modifications on electric conductivity \cite{Hattori:2016cnt,Hattori:2016lqx}, heavy quark dynamics  \cite{Fukushima:2015wck,Das:2016cwd} and quarkonium production \cite{Yang:2011cz,Machado:2013rta,Alford2013a,Guo:2015nsa}, thermal-photon emission \cite{Tuchin:2010gx,Basar:2012bp,Muller:2013ila}, and anomalous effects \cite{Kharzeev:2007jp,Fukushima:2008xe,Kharzeev:2010gd}.

Although the QGP is believed to be dominated by gluons in early stages of heavy ion collisions, it is still tentative to investigate the collective behavior modified by magnetic fields in the framework of magneto-hydrodynamics. Nonetheless, before proceeding to numerical simulations to mimic practical conditions in heavy ion collisions, it is instructive to study the simplified setup with analytic solutions. These simplified models may provide more intuitive pictures or physical insights of the hydrodynamic evolution in simulations. Such simplified solutions in the absence of magnetic fields, which capture qualitative features of heavy ion collisions, have been comprehensively studied \cite{Landau:1953gs,Bjorken:1982qr,Gubser:2010ze,Gubser:2010ui,Hatta:2014gga,Hatta:2014gqa,Hatta:2015ldk}. Among these solutions, the simplest one is the Bjorken solution \cite{Bjorken:1982qr}, which delineates the longitudinal expansion with boost invariance. Recently, there have been some novel developments on the transverse magneto-hydrodynamics (magnetic fields are perpendicular to the fluid velocity) related to heavy ion collisions. In \cite{Roy:2015kma}, the authors derived analytic solutions in the presence of time-dependent magnetic fields transverse to the Bjorken expansion. The presence of magnetic fields modify the energy density of the medium. Following \cite{Roy:2015kma}, the magnetization of the medium is incorporated in \cite{Pu:2016ayh}. In \cite{Pu:2016bxy}, the spatial inhomogeneity of the time dependent magnetic field is further considered, which not only alters the energy density but also the fluid velocity along transverse directions.

In this proceeding, we first review the analytic solution found in \cite{Roy:2015kma} with solely time dependent magnetic fields in Sec\ref{sec_2}. Subsequently, in Sec.\ref{sec_3}, we shift to the primary focus on the solution with spatial inhomogeneity of magnetic fields obtained in \cite{Pu:2016bxy}. In Sec.\ref{sec_4}, we make short summary and concluding remarks.              

\section{Analytic Solutions with Time Dependent $B$ fields}\label{sec_2}
We consider an inviscid fluid coupled to a magnetic field $B^{\mu}$. In the flat spacetime $\eta_{\mu\nu}=\text{diag}\{-,+,+,+\}$, the general form of the energy-momentum tensor is given by \cite{Huang:2009ue}
\begin{equation}
	T^{\mu\nu}=(\epsilon+p+B^{2})u^{\mu}u^{\nu}+(p+\frac{1}{2}B^{2})g^{\mu\nu}-B^{\mu}B^{\nu},\label{eq:EMT_01}
	\end{equation}
where 
	\begin{equation}
	B^{2}=B^{\mu}B_{\mu},\;B^{\mu}=\frac{1}{2}\epsilon^{\mu\nu\alpha\beta}u_{\nu}F_{\alpha\beta}.
	\end{equation}
Here $u^{\mu}$, $\epsilon$, and $p$ correspond to the four velocity of fluid, energy density, and pressure, respectively. Also, $\epsilon^{0123}=-\epsilon_{0123}=1$ represents the Levi-Civita tensor. In our convention, the velocity of the fluid satisfies $u^{\mu}u_{\mu}=-1$. The energy-momentum tensor should follow the conservation equations $\nabla_{\mu}T^{\mu\nu}=0$. In general, the presence of external fields may induce internal electromagnetic fields of the fluid, where the latter are dictated by Maxwell's equations. One should thus solve the conservation equations and Maxwell's equations coupled to each other. In this work, we only focus on the effects of an external magnetic fields and discard the back-reaction from the internal fields. Since the external magnetic field is generated by external sources, it can take an arbitrary form. Therefore, the energy-momentum tensor will be solely governed by the conservation equations. By implementing the projection of $\nabla_{\mu}T^{\mu\nu}=0$ along the longitudinal and transverse directions with respect to $u^{\mu}$, one can rewrite the conservation equations as  
	\begin{eqnarray}\label{cons_original}\nonumber
	u_{\nu}\nabla_{\mu}T^{\mu\nu}\nonumber & = & -(u\cdot\nabla)(\epsilon+\frac{1}{2}B^{2})-(\epsilon+p+B^{2})(\nabla\cdot u)-u_{\nu}\nabla_{\mu}(B^{\mu}B^{\nu})=0,
	\\
	\Delta_{\nu\alpha}\nabla_{\mu}T^{\mu\nu}&=& (\epsilon+p+B^{2})(u\cdot\nabla)u_{\alpha}+\Delta_{\nu\alpha}\nabla^{\nu}(p+\frac{1}{2}B^{2})-\Delta_{\nu\alpha}\nabla_{\mu}(B^{\mu}B^{\nu})=0,
	\end{eqnarray}
where $\Delta^{\nu\alpha}=\eta^{\nu\alpha}+u^{\nu}u^{\alpha}$.

Now, to mimic the condition in heavy ion collisions, we assign the time- dependent magnetic field pointing along $y$ direction, ${\bf B}=B_y(\tau){\hat y}$, and work in Milne coordinates,
\begin{eqnarray}
(\tau,x,y,\eta)=\left(\sqrt{t^2-z^2},x,y,\frac{1}{2}\ln\left(\frac{t+z}{t-z}\right)\right),
\end{eqnarray}  
where $z$ corresponds to the beam direction. For simplicity, we will consider a conformal fluid such that $p=\epsilon/3$. We may now work in the local rest frame $u_{\mu}=(1,{\bf 0})$ and taking $\epsilon=\epsilon(\tau)$ in light of the Bjorken expansion. In the following computations, we will normalize $\tau$ implicitly by an initial time $\tau_0$. It is found in \cite{Roy:2015kma} the second equation of (\ref{cons_original}) vanishes and the first one takes the form
\begin{eqnarray}
\partial_{\tau}\epsilon+\frac{4\epsilon}{3\tau}+B_y\partial_{\tau}B_y+\frac{B_y^2}{\tau}=0,
\end{eqnarray}	
which yields an analytic solution for arbitrary $B_y(\tau)$,
\begin{eqnarray}
\epsilon(\tau)=\frac{1}{\tau^{4/3}}\left(\epsilon_c
-\int^{\tau}_1 du u^{1/3}\left(B_y^2+u B_y\partial_{u}B_y\right)\right),
\end{eqnarray}
where $\epsilon_c$ is determined by initial conditions. As a simple approximation of the time-varying magnetic field in heavy ion collisions, one may consider the power-law decay, $B_y(\tau)=B_0\tau^{n/2}$ with $n<0$. The energy density now becomes
\begin{eqnarray}\nonumber
&&\epsilon(\tau)=\frac{\epsilon_c}{\tau^{4/3}}-\frac{3 B_0^2 (2+n) \tau^n}{8+6 n},
\quad\text{for}\quad n\neq-4/3,
\\
&& 
 \epsilon(\tau)=\frac{\epsilon_c}{\tau^{4/3}}-\frac{B_0^2\log\tau}{3\tau^{4/3}}
 \quad\text{for}\quad n=-4/3.
\end{eqnarray}   
From the above solution, one finds that the energy density can be increased or decreased depending on the exponent $n$. Furthermore, there exists
a special case for $n=2$, in which the correction from magnetic fields vanishes. As discussed in \cite{Roy:2015kma}, this particular scenario occurs due to the "frozen flux theorem". When the magnetic field drops at the same speed as the entropy density $s$ such that
\begin{eqnarray}
(u\cdot\nabla)\left(\frac{B^{\mu}}{s}\right)=\frac{1}{s}\left[(B\cdot\nabla)u^{\mu}+u^{\mu}\nabla\cdot B\right]=0,
\end{eqnarray}
the medium does not "feel" the presence of the magnetic field.

\section{Spatial Inhomogeneity and Transverse Flow}\label{sec_3} 
In the section, we move one step forward to further introduce spatial inhomogeneity of the time-decaying magnetic field, where analytic solutions can be found in the weak-field limit \cite{Pu:2016bxy}. In \cite{Pu:2016bxy}, the magnetic field is assumed to take the form $B_y=\lambda B_s(x)\tau^{n/2}$ with spatial dependence on another transverse direction perpendicular to the magnetic field.   
Now, when $B_y$ has spatial dependence, the transverse fluid velocity should be modified as well due to the spatial gradients of energy density led by magnetic fields. Therefore, one has to solve coupled partial differential equations in (\ref{cons_original}). To simplify the task, the perturbative approach is considered for weak fields, where $\lambda$ can be considered as an expansion parameter. The ansatz of the perturbative solution is given by
\begin{eqnarray}
{\bf B}=\lambda B_s(x)\tau^{n/2}\hat{y}, \quad \epsilon=\epsilon_0(\tau)+\lambda^2 \epsilon_1(\tau,x),\quad u_{\mu}=(1,\lambda^2 u_{x}(\tau,x),0,0),
\end{eqnarray}
where $\epsilon_0(\tau)=\epsilon_c/\tau^{4/3}$. This perturbative solutions has to satisfy the constraint $B^2/\epsilon_0\ll 1$.
Up to $\mathcal{O}(\lambda^2)$, the two differential equations in (\ref{cons_original}) become
\begin{eqnarray}\nonumber\label{two_cons_xdep}
&&
\partial_{\tau}\epsilon_1+\frac{4\epsilon_1}{3\tau}-\frac{4\epsilon_c\partial_{x}u_{x}}{3\tau^{4/3}}+B_y\partial_{\tau}B_y+\frac{B_y^2}{\tau}=0,
\\
&&
\partial_{x}\epsilon_1-\frac{4\epsilon_c\partial_{\tau}u_{x}}{\tau^{4/3}}
+\frac{4\epsilon_cu_{x}}{3\tau^{7/3}}+3B_y\partial_{x}B_y=0.
\end{eqnarray}
The combination of two equations above yields a partial differential equation solely depending on $u_{x}$,   
\begin{eqnarray}\label{equ_xdep}
\tau^2\partial_{x}^2 u_{x}-u_{x}-3\tau^2\partial^2_{\tau}u_{x}+\tau\partial_{\tau}u_{x}+
\frac{3\tau^{7/3}}{4\epsilon_c}\partial_{x}\left(B_y^2+\tau\partial_{\tau}B_y^2\right)=0.
\end{eqnarray}

However, in order to obtain an analytic solution, another approximation is applied in \cite{Pu:2016bxy}. One may now approximate $B_s(x)$ by Fourier series 
\begin{eqnarray}
B^2_y(\tau,x)=\sum_k \tilde{B}_k^2(\tau)\cos(kx), 
\end{eqnarray}
where $k\geq 0$ are now real integers. This approximation converts the task of solving a partial differential equation into solving ordinary differential equations with different moments $k$. Accordingly, one makes the following ansatz,
\begin{eqnarray}
u_{x}(\tau,x)=\sum_m \left(a_m(\tau)\cos(m x)+b_m(\tau)\sin(m x)\right),
\end{eqnarray}
and solve (\ref{equ_xdep}). For each moment with $k>0$, we find $m=k$ and $a_m(\tau)=0$, while $b_k(\tau)$ is solved from the following ordinary differential equation,
\begin{eqnarray}\label{eq_bk}
(3\tau^2\partial_{\tau}^2-\tau\partial_{\tau}+k^2\tau^2+1)b_k(\tau)+\frac{3B_k^2}{4\epsilon_c}k(n+1)\tau^{n+7/3}=0.
\end{eqnarray}
As shown in \cite{Pu:2016bxy}, the solution for each moment takes an complicated form but has an analytic expression. In addition, the integration constants are fixed by the asymptotic solutions in late times with vanishing magnetic fields. Given $b_k$, one can also acquire the energy density through (\ref{two_cons_xdep}) 
\begin{eqnarray}\label{solve_e1}
\epsilon_1(\tau,x)=-\frac{3 B_0^2 (2+n) \tau^n}{8+6 n}
-\sum_{k\neq 0}\frac{\cos(kx)}{k}\left(\frac{4\epsilon_c\partial_{\tau}b_k(\tau)}{\tau^{4/3}}-\frac{4\epsilon_cb_k(\tau)}{3\tau^{7/3}}+\frac{3k}{2}\tilde{B}^2_k
(\tau)\right).
\end{eqnarray}  
  
\begin{figure*}[ht!]
	\centering
	\begin{tabular}{cc}
		\includegraphics[width=0.48\linewidth]{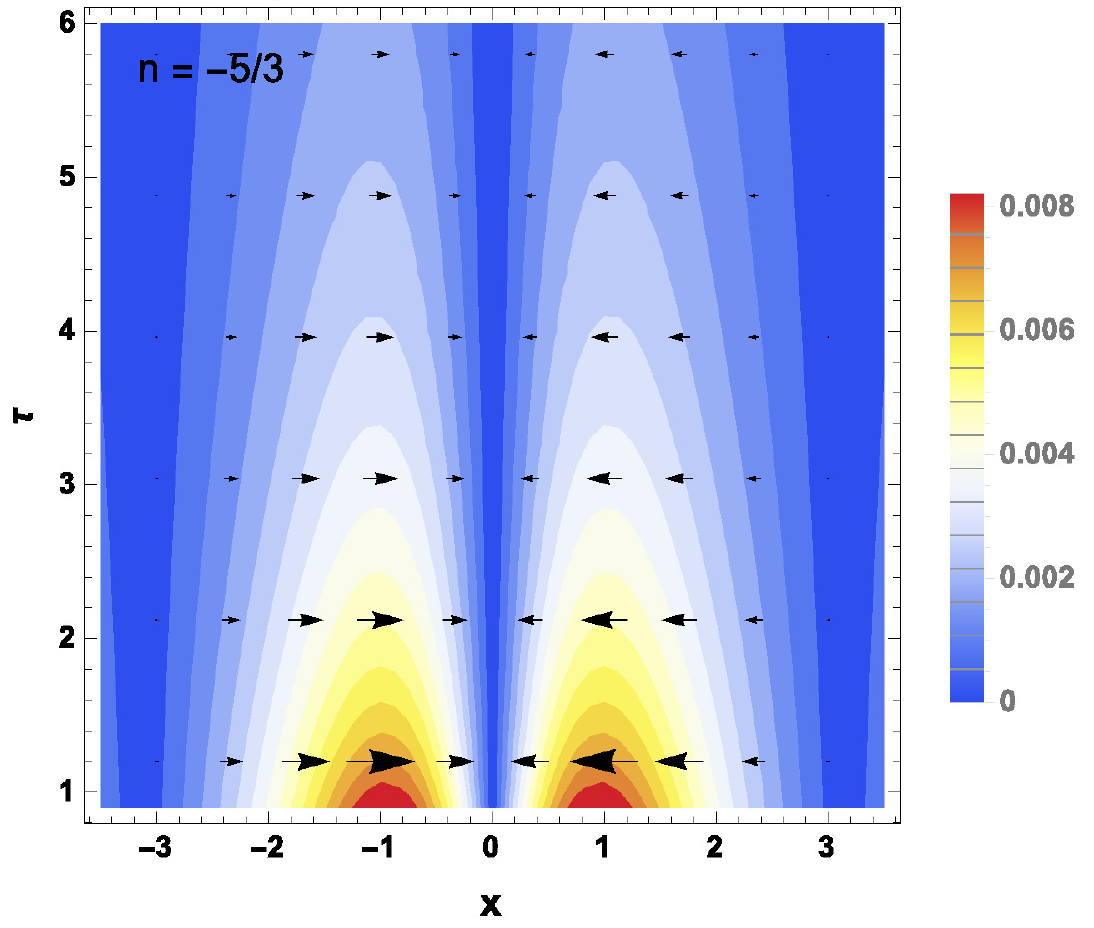} &
		\includegraphics[width=0.48\linewidth]{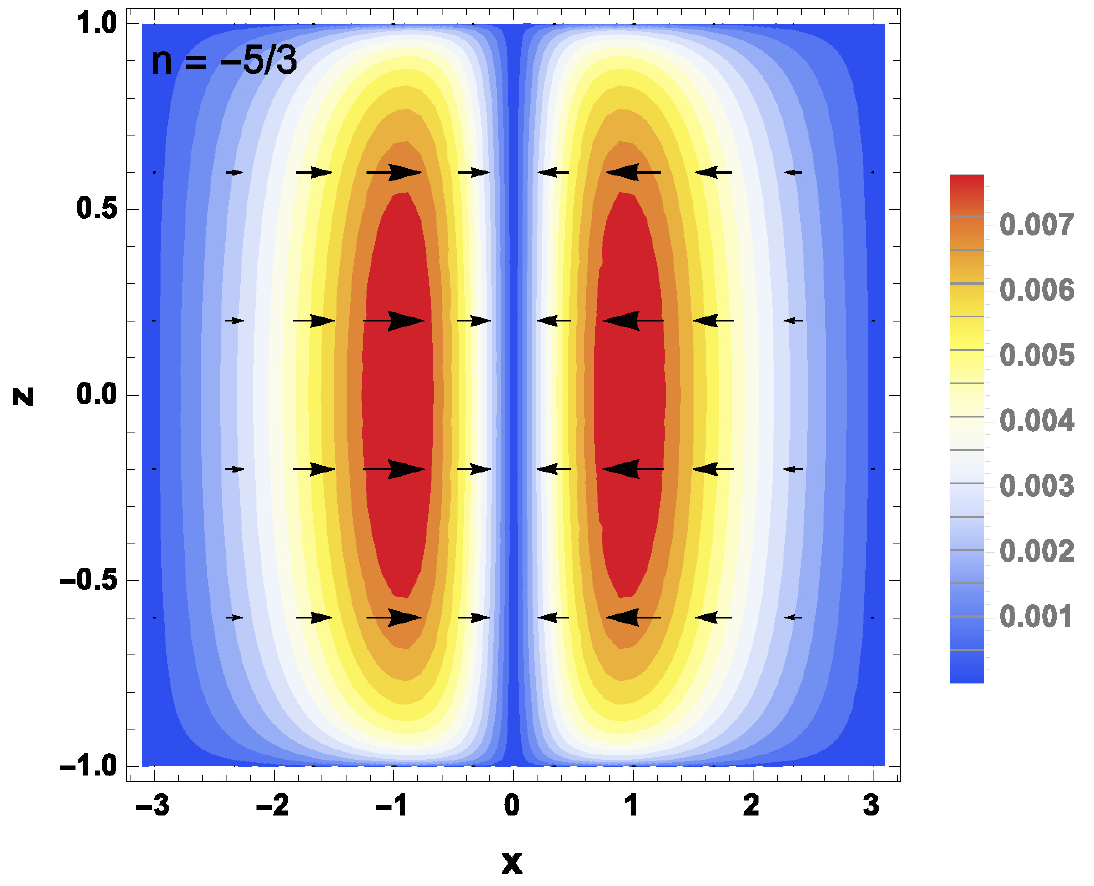}
	\end{tabular}
	\caption{Transverse velocity plot for $v_x(\tau,x)=u^x/u^{\tau}$ with $n=-5/3$ \cite{Pu:2016bxy}.}
	\label{Fig1}
\end{figure*}

\begin{figure*}[ht!]
	\centering
	\begin{tabular}{cc}
		\includegraphics[width=0.48\linewidth]{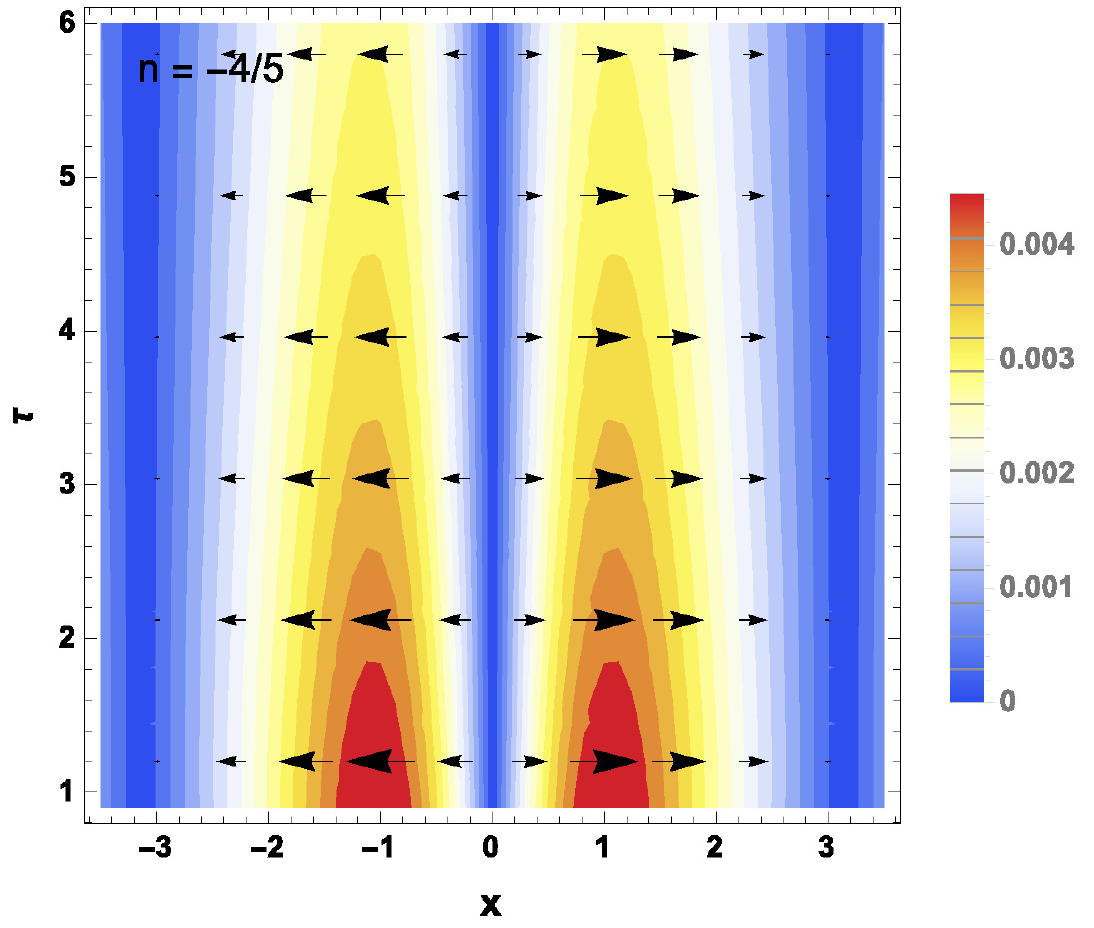} &
		\includegraphics[width=0.48\linewidth]{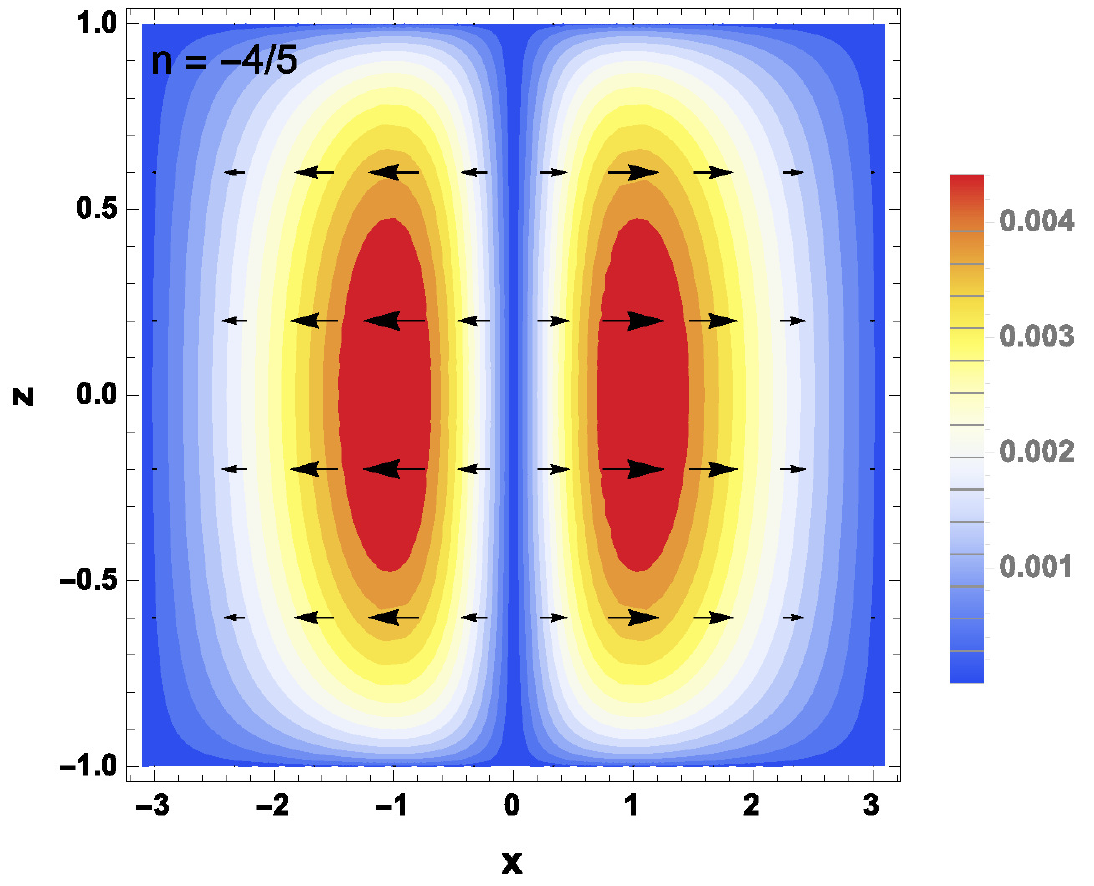}
	\end{tabular}
	\caption{Transverse velocity plot for $v_x(\tau,x)=u^x/u^{\tau}$ with $n=-4/5$ \cite{Pu:2016bxy}.}
	\label{Fig2}
\end{figure*}

Although the general solution is too complicated to show some enlightening features, it reduces to a simple solution when $n=-1$, which reads
\begin{eqnarray}
u_x=0,\quad \epsilon_1=-\frac{3\tilde{B}^2_y(x)}{2\tau},
\end{eqnarray} 
for arbitrary $\tilde{B}_y(x)$. The vanishing $u_x$ at a particular exponent then implies the change of directions of transverse velocity led by magnetic fields with distinct exponents. As a concrete example, the Gaussian profile of the magnetic field is considered
\begin{eqnarray}
{\bf B}=B_y(\tau,x)\hat{y}=B_c\tau^{n/2}e^{-x^2/2}\hat{y}.
\end{eqnarray}
The magnetic field is then decomposed into Fourier series with the truncation up to the forth moment. Although the periodic properties of cosine series yields deviations at the fringes of the Gaussian distribution, such an approximation is sufficient for the central region where the magnetic field is most prominent. In Fig.\ref{Fig1} and Fig.\ref{Fig2}, the flow profiles led by magnetic fields with different decay exponents are presented. As anticipated, the flow decreases with respect to time. Also, it is found that the fluid velocities indeed have opposite orientations for $n<-1$ and $n>-1$.     
  
It is argued in \cite{Pu:2016bxy} that the change of directions of $v_x$ with distinct values of $n$ stems from the conservation of magnetic flux. To simplify the conditions, one may consider two extreme cases, which correspond to $n\ll -1 $ and $n\gg -1$. For $n\ll -1$, the time scale of the magnetic field is much shorter than the one for the expanding medium, one thus approximates such a condition as a static medium in the presence of a time-decreasing magnetic field $B_y(t)$ with a Gaussian distribution in $x$. The total magnetic flux going through the medium now drops with respect to time. The medium is thus pushed inward to the central region $x=0$ in order to preserve the flux. On the contrary, for $n\gg -1$, the magnetic field decays much slower than the expansion of the medium. One thus approximates the situation with the presence of a static magnetic field $B_y(x)$ as a Gaussian function of $x$ in a medium expanding along the $z$ direction. In such a case, the total magnetic flux of the medium increases with respect to time. To reduce the flux, the medium hence expands along the $\pm x$ directions. The case for $n=-1$ may correspond to the situation in which the magnetic flux is balanced by the expansion of the medium and the decrease of the magnetic field, which thus results in the absence of transverse flow. 

\section{Summary and Concluding Remarks}\label{sec_4}
In this proceeding, we review the analytic solutions in ideal and transverse magneto-hydrodynamics under Bjorken expansion with spacetime-dependent magnetic fields. Particularly, the spatial inhomogeneity of magnetic fields further induces transverse flow. Nevertheless, the impact of such a scenario on hadronic flow in heavy ion collisions requires more pragmatic studies. Recently, there have been some numerical studies of transverse flow in magneto-hydrodynamics \cite{Pang:2016yuh,Inghirami:2016iru}. However, due to different theoretical setup and uncertainties of the spacetime profile of magnetic fields in heavy ion collisions, the influence from magnetic fields on collective behaviors of QGP is still inconclusive. On the other hand, since magnetic fields only interact with quarks, it is rather challenging but necessary to separate degrees of freedom of quarks and gluons properly in collective motions and search for suitable observables to isolate the magnetic-field induced effect.    

Acknowledgments: S.P. is supported by the Alexander von Humboldt Foundation, Germany and D.Y. is supported by the RIKEN
Foreign Postdoctoral Researcher program.       	

\end{document}